\documentclass[aps,prc,twocolumn,showpacs,floatfix]{revtex4}   
\usepackage{graphicx}
\begin{document}

\title{Hindrance in the fusion of $^{48}$Ca+$^{48}$Ca} 
\author{H. Esbensen$^1$, C. L. Jiang$^1$ and A. M. Stefanini$^2$}
\affiliation{$^1$ Physics Division, Argonne National Laboratory, Argonne, Illinois 60439}
\affiliation{$^2$ INFN, Laboratori Nazionali di Legnaro, I-35020 Legnaro (Padova), Italy}
\date{\today}
\begin{abstract}
The coupled-channels technique is applied to analyze recent fusion 
data for $^{48}$Ca+$^{48}$Ca. The calculations include the excitations
of the low-lying $2^+$, $3^-$ and $5^-$ states in projectile and target,
and the influence of mutual excitations as well as the two-phonon 
quadrupole excitations is also investigated.
The ion-ion potential is obtained by double-folding the nuclear 
densities of the reacting nuclei with the M3Y+repulsion effective 
interaction but a standard Woods-Saxon potential is also applied.
The data exhibit a strong hindrance at low energy compared to calculations 
that are based on a standard Woods-Saxon potential but they
can be reproduced quite well by applying the M3Y+repulsion 
potential with an adjusted radius of the nuclear density.
The influence of the polarization of high-lying states
on the extracted radius is discussed.
\end{abstract}
\maketitle

\section{Introduction}

Heavy-ion fusion reactions are sensitive probes of the nuclear surface 
of the reacting nuclei. Roughly speaking, the height of the Coulomb 
barrier is determined by the radii and diffuseness of the densities, 
and the enhancement of fusion at subbarrier energies is governed by 
couplings to the excitation of low-lying surface modes \cite{baha}.
This picture may not always succeed in reproducing the measured
cross sections. In practical coupled-channels calculations it is
often necessary to make adjustments, either in the structure input
or in the ion-ion potential. The adjustments may reflect the influence 
of high-lying states or other reaction channels that are not treated 
explicitly in the calculations. Another complication is the hindrance
of fusion which occurs at low energies and very small cross sections
\cite{jiangsys}. The hindrance can be explained, for example, by using 
a shallow potential in entrance channel \cite{misi75} or by modeling 
the fusion dynamics for touching and overlapping nuclei \cite{ichikawa}.
 
In this work the coupled-channels technique is applied to analyze 
the fusion data for $^{48}$Ca+$^{48}$Ca that were recently measured down to 
very small cross sections below 1 $\mu$b \cite{stef4848}.
An analysis of the data provides the opportunity to investigate 
whether the fusion hindrance, which is a well established phenomenon 
in extreme subbarrier fusion reactions of medium-heavy systems with 
large negative Q-values \cite{jiangsys}, also occurs in the fusion 
of calcium isotopes with near zero Q-values.
An indication of a hindrance in the fusion of $^{48}$Ca+$^{48}$Ca 
has already been observed \cite{stef4848} because the low-energy 
data could only be reproduced by coupled-channels calculations that 
employ a very large diffuseness of the ion-ion potential.
However, the hindrance observed there was not strong enough to 
show an $S$ factor maximum in the energy region of the measurement.

The ion-ion potential and the nuclear couplings that will be used 
are derived as in previous work \cite{misi75} from the double-folding
of the densities of the reacting nuclei and the M3Y+repulsion effective 
interaction.  
Since the low-lying structure of $^{48}$Ca is fairly well established and
the influence of transfer reactions is always suppressed for symmetric 
systems, it is expected that the simple picture of fusion described 
above would apply to the fusion of $^{48}$Ca+$^{48}$Ca. 
On the other hand, it is well known that the excitation and/or
polarization of high-lying states that are included explicitly 
in the coupled-channels calculations can lead to a negative energy 
shift of the calculated fusion cross sections \cite{takigawa}.
This is effectively equivalent to increasing the radii of the 
reacting nuclei. The radius that is extracted by optimizing the fit 
to the fusion data can therefore be too large because it can be 
contaminated by the polarization of states which are not included 
explicitly in the calculations. 
It is of interest to see how the extracted radius depends on the 
model space of excited states that are included in the calculations, 
and how well it compares to the expected matter radius of $^{48}$Ca.

The study of the fusion of different calcium isotopes started more than
twenty years ago with the measurements by Aljuwair et al. \cite{aljuwair} 
but the cross sections were only measured down to about 1 mb. 
The most challenging theoretical issue at that time was to explain the 
fusion of $^{40}$Ca+$^{48}$Ca which appeared to be strongly enhanced 
by couplings to transfer reactions, in particular to those with positive 
Q values \cite{landowne,fricke}. A strong motivation for reviving
the study of the fusion of calcium isotopes is that, in addition to 
$^{48}$Ca+$^{48}$Ca \cite{stef4848}, the fusion of $^{40}$Ca+$^{48}$Ca
has also recently been measured down to the 1 $\mu$b \cite{jiang4048},
and a new measurement for $^{40}$Ca+$^{40}$Ca is underway \cite{stef4040}.
In order to be able to focus on and isolate the effect of transfer 
on the fusion of the asymmetric $^{40}$Ca+$^{48}$C system, it is 
necessary first to develop a good description of the fusion of the 
two symmetric systems, 
and the present work is a step in that direction.

The nuclear structure properties of $^{48}$Ca are discussed in the next 
section and Sect. III describes the construction of the ion-ion potential. 
The coupled-channels technique is summarized in Sect. IV together with
an analysis of the data that is based on Woods-Saxon potentials.
The analysis based on the M3Y+repulsion potential is presented in Sect. V.
Finally, the conclusions are given in Sect. VI.

\section{Nuclear structure input} 

The nuclear structure input to the coupled-channels calculations
is shown in Table \ref{structure}. The elastic channel and the 
one-phonon excitations of the low-lying $2^+$, $3^-$ and $5^-$ states
in projectile and target results in a total of 7 channels..
The coupling strengths for the excitation of these states 
are taken from Ref. \cite{flem} where they were calibrated by analyzing the 
elastic and inelastic scattering of $^{16}$O on calcium isotopes \cite{rehm}. 
It should be noted that the Coulomb and nuclear coupling strengths, 
expressed in Table \ref{structure} by the values of $\beta R/\sqrt{4\pi}$, 
are different. The Coulomb couplings are in most cases consistent with 
the currently adopted electromagnetic transition probabilities or 
$B$-values \cite{ENDSF} that are quoted in the third column of 
Table \ref{structure}.

Also shown in Table \ref{structure}  are the $0^+$, $2^+$ and $4^+$ members 
of two-phonon quadrupole excitation. 
The adopted $B(E2)$-values \cite{ENDSF} shown in the third column 
of the second part of Table \ref{structure} can be combined 
into an effective two-phonon excitation.  
For example, the effective $B(E2)$ value for the two- to one-phonon 
transition is given by the sum,
\begin{equation}
B(E2,{\rm 2ph}\rightarrow{\rm 1ph}) = \sum_{I=0,2,4}
\langle 2020|I0\rangle^2 \ B(E2,I\rightarrow 2),
\end{equation}
and the two-phonon excitation energy is obtained as the weighted average 
of the individual two-phonon excitations energies \cite{prc72}.
The parameters obtained for the effective two-phonon quadrupole 
excitation are shown in the last line of Table \ref{structure}.  
Including the effective two-phonon quadrupole  excitations in the 
coupled-channels calculations, in addition to the 7 channels 
mentioned above, leads to a total of 9 channels. 
Unfortunately, nothing is known about the two-phonon 
excitations of the $3^-$ and $5^-$ states so they will be ignored.

To get a feeling of the influence of higher-order excitations on the 
calculated fusion cross sections one can also include all of the 15 mutual 
excitations channels that are generated from the six one-phonon 
excitations presented in Table \ref{structure}. Together with the 
basic 9 channels mentioned above, that sums up to a total of 24 channels.
This will be referred to as the full calculation and the results will be 
compared to the fusion data and to calculations that include the 9 channels 
described above, as well as the no-coupling limit in which case there
is only 1 channel.

\begin{table}
\caption{Nuclear structure input for $^{48}$Ca. 
Values marked  with * are from Ref. \cite{flem}.
The $B(E\lambda)$ values are from Ref, \cite{ENDSF}. They are consistent 
with the Coulomb excitation parameters $(\beta R)_C/\sqrt{4\pi}$, 
except for the $3^-$ state where Ref. \cite{flem} uses a larger value, 
$B(E3)$ = 6.8 W.u. The second part of the Table shows the 3 
transitions that determine the effective two-phonon quadrupole state. 
Its excitation energy and couplings to the $2^+_1$ state are shown in 
the last line.}
\label{structure}
\begin{tabular} {|c|c|c|c|c|}
\colrule
 $I^\pi$ & $E_x$ (MeV) & $B(E\lambda)$ (W.u.) &
$\frac{(\beta R)_C}{\sqrt{4\pi}}$ (fm) &
$\frac{(\beta R)_N}{\sqrt{4\pi}}$ (fm) \\
\colrule
 $2^+_1$  & 3.832 & 1.71(9)  & 0.126*& 0.190* \\
 $3^-$         & 4.507 &  5.0(8)  & 0.250*& 0.190* \\
 $5^-$         & 5.146 &          & 0.049*& 0.038* \\
\colrule
 $0^+_2\rightarrow 2^+_1$ & 4.283 & 10.1(6)  &[0.098]&       \\
 $4^+_1\rightarrow 2^+_1$ & 4.503 & 0.261(6) &[0.025]&       \\
 $2_2^+\rightarrow 2_1^+$ & 5.311 &  9(9)    &[0.111]&       \\
\colrule
 Eff 2PH       & 4.849 & 4.7(29)  & 0.15 & 0.15 \\
\colrule
\end{tabular}
\end{table}

\section{The ion-ion potential}


The parameters of the Woods-Saxon (WS) potentials that will be used 
in this work,
\begin{equation}
U_{\rm WS}(r) = \frac{-V_0}{1+\exp((r-R_0)/a)},
\end{equation}
are those proposed in Ref. \cite{BW}, Eqs. (III.2.40-45).
The parameters for the system $^{48}$Ca+$^{48}$Ca are $a$ = 0.662 fm 
for the diffuseness and $V_0$ = 64.10 MeV for the depth of the potential.
We refer to this potential as the `standard' WS potential because
its diffuseness is consistent with the analysis of elastic scattering 
data \cite{BW}.
The radius $R_0$, on the other hand, will be treated as a free parameter 
and it will be adjusted to optimize the fit to the data in the 
coupled-channels calculations that are discussed in the next section. 

There is an interesting point concerning the isospin dependence of the
nuclear potential proposed in Ref. \cite{BW}. It enters through the 
average nuclear surface tension $\gamma$ of the reacting nuclei,
\begin{equation}
\gamma = 0.95 \Bigl[1-1.8 \ \frac{N_a-Z_a}{A_a} \ \frac{N_b-Z_b}{A_b}\Bigr]
\ \ {\rm MeV \ fm^{-2}},
\label{gamma}
\end{equation}
according to Ref. \cite{BW}, Eq. (III.2.30). The correction factor 
in Eq. (\ref{gamma}), which depends on neutron excess,
is equal to one in reactions that involve $^{40}$Ca. In the case of
$^{48}$Ca+$^{48}$Ca, the correction factor reduces the surface tension
by 5\%; this correction is included in the value of the
depth parameter $V_0$ mentioned above.


The M3Y+repulsion potential is calculated using the double-folding 
technique described in Ref. \cite{misi75}. It is based on the Reid 
parametrization of the M3Y effective nucleon-nulceon interaction
\cite{bertsch}.
The spherical nuclear densities are parametrized by 
\begin{equation}
\rho(r) = 
\frac{1}{2} \ \frac{\rho_0 \exp(R/a)}{\cosh(r/a)+\cosh(R/a)},
\label{rho}
\end{equation}
where $R$ and $a$ are the radius and diffuseness parameters, respectively,
and $\rho_0$ is a normalization constant. It is seen that the density,
Eq. (\ref{rho}), approaches the Fermi function density 
$\rho_0/(1+\exp((r-R)/a))$ for $R/a>>1$. 
The parametrization Eq. (\ref{rho}) is used here because it 
has some very useful analytic properties as discussed in the Appendix 
of Ref. \cite{misopb}. For example, the Fourier transform has an 
analytic form which simplifies the calculation of the double-folding
potential in Fourier space from the expression \cite{misi75},
\begin{equation}
U(r) = \frac{1}{2\pi^2} \ 
\int k^2dk \ \rho(k) \ \rho(k) \ v_{nn}(k) \ j_0(kr),
\end{equation}
where $v_{nn}(k)$ is the Fourier transform of the effective 
nucleon-nucleon interaction, and $j_0(x)=\sin(x)/x$
is a spherical Bessel function. Another advantage of the 
parametrization Eq. (\ref{rho}) is that the mean-square 
radius is given by the simple expression
\begin{equation}
\langle r^2 \rangle = \frac{3}{5} \
\Bigl( R^2 + \frac{7}{3} \ (a\pi)^2\Bigr).
\label{rms}
\end{equation}

The repulsive part of the M3Y+repulsion potential is determined by two 
parameters, namely, the strength $v_r$ of the contact effective interaction,
\begin{equation}
v_{nn}^{\rm rep}({\bf r}) = v_r \ \delta({\bf r}),
\end{equation}
that generates it, and the diffuseness $a_r$ of the densities that are
applied in the double-folding calculation \cite{misi75}.
The radius parameter $R$ of the densities, on the other hand, is kept the 
same as in the calculation of the direct and the exchange parts of the M3Y 
double-folding potential. The diffuseness of the density that is 
used in calculating the direct and the exchange part of the M3Y potential
is kept fixed with the value $a$ = 0.54 fm. 

The two parameters $a_r$ and $v_r$ of the repulsive part of the potential 
are constrained so that total nuclear potential energy, $U_N(r)$, for 
completely overlapping nuclei is consistent with the equation-of-state. 
That leads to the relation \cite{misi75}, 
\begin{equation}
U_N(r=0) \approx \frac{A_p}{9} \ K,
\label{inkom}
\end{equation}
where $A_p$ is the mass number of the smaller nucleus and $K$ is the
nuclear incompressibility. For $^{48}$Ca+$^{48}$Ca the value $K$ = 223.7 
MeV predicted by the Thomas-Fermi model of Myers and \' Swi\c atecki 
\cite{myers} will be used. 
Thus there are essentially only two free parameters of the M3Y+repulsion
interaction, namely, the radius $R$ and the diffuseness parameter $a_r$.
They will be adjusted to optimize the fit to the fusion data. 
The strength $v_r$ of the repulsive interaction, on the other hand, is 
constrained for given values of $R$ and $a_r$ by the 
nuclear incompressibility according to Eq. (\ref{inkom}).


\begin{figure}
\includegraphics[width = 8cm]{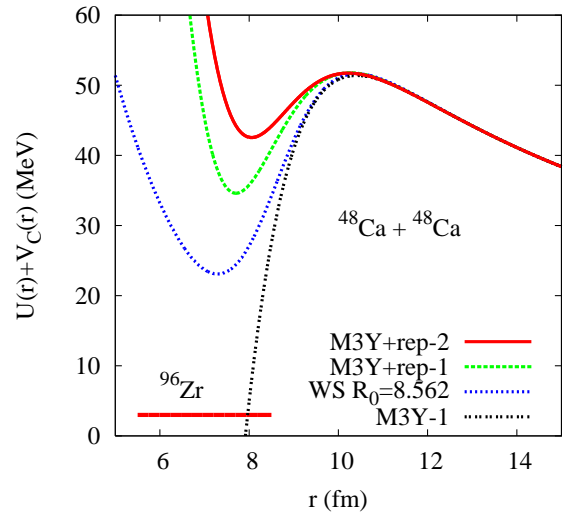}
\caption{\label{4848pot} (Color online)
Entrance channel potentials for $^{48}$Ca+$^{48}$Ca obtained from
the Woods-Saxon (WS, with $R_0$ = 8.562 fm) 
and the pure M3Y (M3Y-1) potentials. 
The two upper curves are the shallow M3Y+repulsion potentials
determined in Sect. V.  The energy of ground state of the compound 
nucleus $^{96}$Zr is also indicated.}
\end{figure}

Some of the entrance channel potentials 
that are used in this work are illustrated in Fig. \ref{4848pot}.
The height of the Coulomb barrier is essentially the same for all four
potentials but the thickness of the barrier is very different. 
The (blue) dashed curve is the entrance channel potential for the 
Woods-Saxon potential. 
It has the minimum pocket energy $V_{min}$ =23.09 MeV and the height 
of the Coulomb barrier is $V_{CB}$ = 51.63 MeV.  The latter potential
was determined by 
optimizing the fit to the fusion data with center-of-mass energy
larger than 50 MeV. This energy cut was chosen because the fusion 
hindrance phenomenon sets in below 50 MeV as we shall see in 
the next section.

The upper two curves in Fig. \ref{4848pot} are the M3Y+repulsion 
entrance channel potentials that are obtained in Sect. V. 
They were determined by optimizing the fit to the fusion data in 
coupled-channels calculations that include the 24 channels described 
in Sect. II. 
There are two solutions, the M3Y+rep-1 and M3Y+rep-2 potentials, 
which are discussed in detail in Sect. V.
These potentials are shallower than the standard Woods-Saxon 
potential, which is a characteristic feature of the 
M3Y+repulsion potentials that have been extracted from fusion 
data \cite{misi75}.  

Finally, the entrance channel potential for the pure M3Y(+exchange) 
potential is also shown. It is unrealistic because it is deeper 
than the ground state energy of the compound nucleus $^{96}$Zr 
which is indicated by the thick horizontal line.


\section{Coupled-channels calculations}

The coupled-channels calculations are performed in the rotating frame 
approximation, and the fusion cross sections are determined by imposing 
ingoing-wave boundary conditions at the position of the minimum of the
pocket in the entrance potential. This procedure is commonly used and 
is described, for example, in Refs. \cite{misi75,oge}. 
In the present work no imaginary potential will be applied. The fusion 
cross section will therefore vanish when the center-of-mass energies is 
lower than the minimum energy of the pocket in the entrance channel potential.

The nuclear potential enters the coupled equations both directly by 
determining the entrance channel potential 
and indirectly by determining the nuclear couplings to first and second 
order in the deformation amplitudes through the first and second 
derivatives of the nuclear potential (see Ref. \cite{oge} for details.) 
There are in principle couplings of even higher order and to higher-lying
states \cite{hagino} but they will be ignored in the present study, partly 
because they are poorly known and partly because they are not expected 
to play a large role in the fusion of the not-so-heavy 
system $^{48}$Ca+$^{48}$Ca. 
This expectation is based on the experience gained in Ref. \cite{prc72}.
However, the polarization of high-lying states \cite{takigawa} that are 
not included in the calculations could distort the analysis.



\begin{figure}
\includegraphics[width = 8cm]{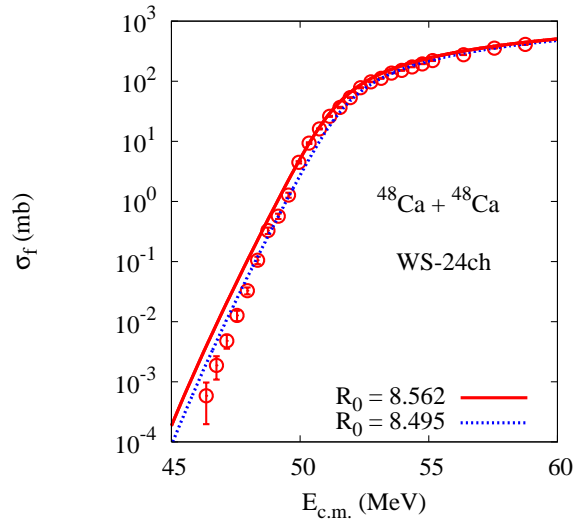}
\caption{\label{4848wsff} (Color online) 
Fusion cross sections for $^{48}$Ca+$^{48}$Ca. 
The error bars reflect the statistical uncertainties.
The curves are coupled-channels calculations with 24 channels 
that use Woods-Saxon (WS) potentials with two different radii.}
\end{figure}

The fusion data for $^{48}$Ca+$^{48}$Ca \cite{stef4848} are compared in
Fig. \ref{4848wsff} to two coupled-channels calculations that 
are based on standard Woods-Saxon potentials \cite{BW} with the 
diffuseness $a$ = 0.662 fm and depth $V_0$ = 64.10 MeV.
All 24 channels described in Sect. II were included in the calculations.
It is seen that the data are hindered at low energies and the energy 
dependence is much steeper than predicted by the calculations.
The dashed curve in Fig. \ref{4848wsff} is the best fit to all data 
points; it is achieved with radius $R_0$ = 8.495 fm but the fit is very 
poor with an average $\chi^2$ per data point of $\chi^2/N$ = 7.9,
including the statistical uncertainties and a systematic error of 7\%. 

The solid curve in Fig. \ref{4848wsff} is based on the slightly larger 
radius, $R_0$ = 8.562 fm. It provides a better account of the data near 
and above the the Coulomb barrier as discussed below. The associated
entrance channel potential is the (blue) dashed curve shown in 
Fig. \ref{4848pot}.


\begin{figure}
\includegraphics[width = 8cm]{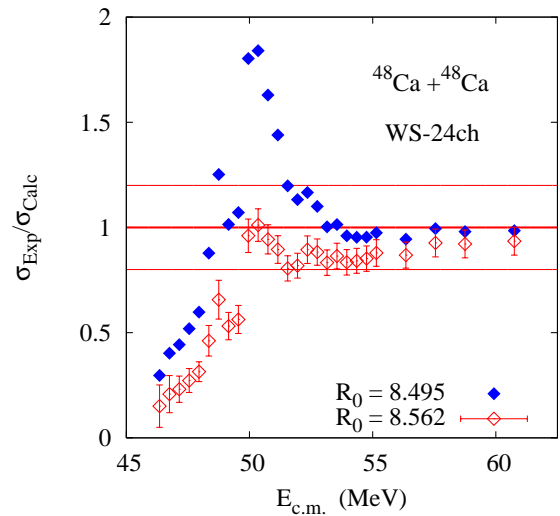}
\caption{\label{4848wsfr} (Color online) 
Ratios of the measured and calculated cross sections shown in Fig. 
\ref{4848wsff}. The error bars on the open diamonds 
were determined by the statistical uncertainties and 
a 7\% systematic error.}
\end{figure}

The behavior of the hindrance in the fusion of $^{48}$Ca+$^{48}$Ca 
is illustrated in Fig. \ref{4848wsfr} in terms of 
the ratio of the measured and calculated fusion cross sections. 
It is seen that the ratio with respect to the best fit to the data
(the solid diamonds) has a strong peak near 50 MeV, slightly below 
the Coulomb barrier which is at $V_{CB}\approx$ 52 MeV.
The ratio drops quickly at energies below 50 MeV. This is attributed
to the fusion hindrance phenomenon. In fact, the steep falloff 
with decreasing energy observed in the comparison to standard 
coupled-channels calculations was the signature that was first used 
to identify the fusion hindrance \cite{jiangniy}. 
Later it was shown that the hindrance is often so strong that
the $S$ factor for fusion develops a maximum at very low energies.
Moreover, it was realized that an $S$ factor maximum together with  
the energy $E_S$ of the maximum is a good quantitative way 
to characterize the fusion hindrance phenomenon \cite{jiang2004}.

Since the fusion hindrance occurs at low energies one may exclude 
the low energy region and focus on reproducing the data at higher energies.
The result of this approach is shown in Fig.  \ref{4848wsff} by the solid 
curve which is based on a slightly larger radius, $R_0$ = 8.562 fm. 
The larger radius implies larger cross sections below the Coulomb barrier 
but that gives a better description of the excitation function in the
barrier region. 
The radius of the Woods-Saxon potential was therefore chosen so that 
the ratio of the measured and calculated cross sections essentially
is a constant above 50 MeV. This is illustrated by the open 
diamonds Fig. \ref{4848wsfr}.
It is seen that the fusion hindrance sets in very strongly below 50 MeV, 
where the ratio falls off very steeply with decreasing energy. 


\section{Analysis based on the M3Y+repulsion potential}

The parameters of the M3Y+repulsion potential that provides the best fit 
to the data were determined using an improved calibration procedure.
For a given nuclear radius parameter $R$ of $^{48}$Ca  
and a given diffuseness $a_r$ of the density used in calculating the 
repulsive part of the potential, the strength of the repulsive term 
$v_r$ was adjusted so that the incompressibility $K$ = 223.7 MeV 
was achieved in Eq. (\ref{inkom}).
Having determined the nuclear potential, coupled-channels calculations 
were performed 
and the average $\chi^2$ per data point, $\chi^2/N$, was calculated
from the statistical uncertainties and a systematic error of 7\%.

\begin{figure}
\includegraphics[width = 8cm]{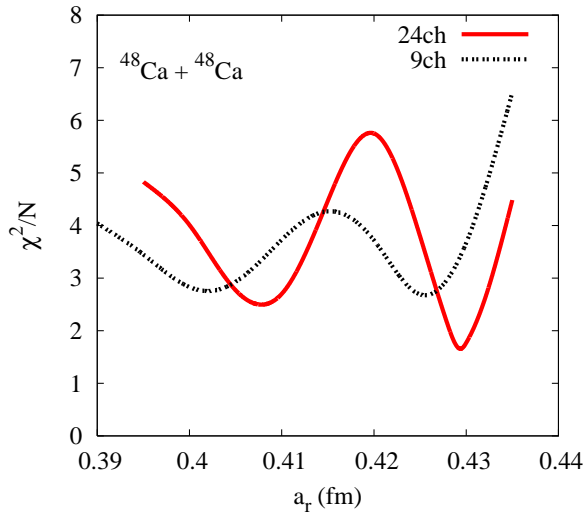}
\caption{\label{xki2} (Color online) Results of the  
$\chi^2$ analysis of the $^{48}$Ca+$^{48}$Ca fusion data \cite{stef4848}. 
The $\chi^2/N$, minimized with respect to the radius $R$, 
is shown as function of the diffuseness parameter $a_r$.
The solid curve (24ch) is for calculations that 
include 24 channels; the dashed curve (9ch) is for 9 channels.}
\end{figure}

The above procedure was repeated with different values of the radius $R$
for a fixed diffuseness parameter $a_r$ until a minimum $\chi^2/N$ was 
found. The whole process was repeated for a new value of $a_r$.
The results of this process are illustrated in Fig. \ref{xki2}
where the $\chi^2/N$, minimized with respect to the radius $R$, 
is plotted as a function of the diffuseness parameter $a_r$. 
The dashed curve is the result of calculations that include the 9 channels.
The solid curve is the result obtained with all 24 channels 
described in Sect. II.

\begin{figure}
\includegraphics[width = 8cm]{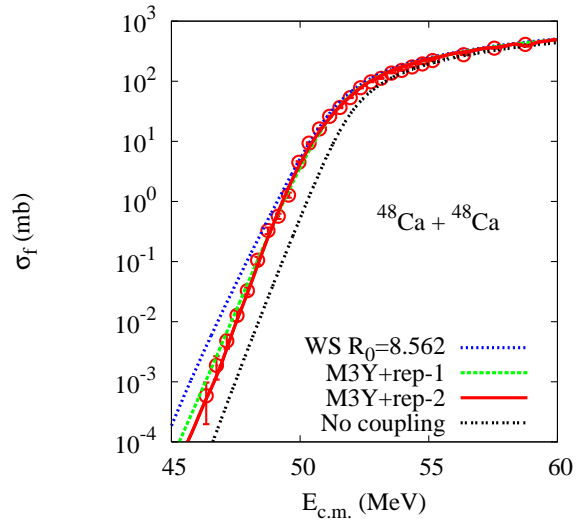}
\caption{\label{m3y4848ff} (Color online) 
Fusion cross sections for $^{48}$Ca+$^{48}$Ca.
The upper two curves are coupled-channels calculations with 24 channels
that are based on the Woods-Saxon (WS) and the M3Y+repulsion
(M3Y1+rep-1 and M3Y+rep-2) potentials discussed in the text. 
The lowest dashed curve is the no-coupling limit based on the 
same M3Y+repulsion potential.}
\end{figure}

The fusion data of Ref. \cite{stef4848} are compared in Fig. 
\ref{m3y4848ff} to various calculations. The solid curve is
the result of coupled-channels calculations associated with 
the deepest minimum in Fig. \ref{xki2} which has a $\chi^2$ 
per data point of 1.66.  
The upper (blue) dashed curve is the cross sections obtained
in similar calculations using the Woods-Saxon potential with 
the radius $R_0$ = 8.562 fm. 
All 24 channels described in Sect. II were included in both
sets of calculations.
The only difference between the two calculations is the 
choice of the nuclear potential, and it is seen that the
shallow M3Y+repulsion potential labeled M3Y+rep-2 is a 
much better choice. 

The lowest dashed curve in Fig. \ref{m3y4848ff} is the cross
section obtained in the no-coupling limit (i.~e., with only 
1 channel) using the same M3Y+repulsion potential that was used
to produce the solid curve. Comparing the two curves, it is seen
that the effect of the couplings to the 24 channels is equivalent 
to shifting the no-coupling limit almost 1 MeV to lower energies.

\subsection{Details of the analysis}

The minima of the curves shown in Fig. \ref{xki2} define the 
stable solutions of the $\chi^2$ analysis of the fusion data since 
they are minima with respect to variations in both $a_r$ and $R$.
There are two local minima for each set of calculations and the 
parameters of the M3Y+repulsion interactions that determine them 
are given in Table \ref{xki4848}. It is seen the two solutions
obtained with 9 coupled have almost the same $\chi^2/N$.
It is not clear what causes the existence of two solutions. 
The main difference between them is that the energy of pocket 
in the entrance channel potential, $V_{min}$, is about 
8 MeV deeper in the solution with the smaller radius $R$.

\begin{table}
\caption{Parameters of the M3Y+repulsion potential associated with the 
$\chi^2$ minima in Fig. \ref{xki2}. Results are shown for coupled-channels 
calculations that include 9 and 24 channels, respectively.
The last three columns show the minimum of the pocket $V_{min}$, the height 
of the Coulomb barrier $V_{CB}$, and the $\chi^2$ per data point.} 
\label{xki4848}
\begin{tabular} {|c|c|c|c|c|c|c|}
\colrule
No. of Ch. & $R$  & $a_r$ & $v_r$        & $V_{min}$ & $V_{CB}$ & $\chi^2/N$ \\
        & (fm) & (fm)  & (MeV fm$^3$) & (MeV)     & (MeV) & \\
\colrule
9  & 3.775 & 0.4025 & 480.1 & 33.58 & 51.67 & 2.76 \\ 
9  & 3.810 & 0.4250 & 504.2 & 41.66 & 51.60 & 2.69 \\ 
24 & 3.745 & 0.4070 & 481.8 & 34.61 & 51.77 & 2.52 \\ 
24 & 3.798 & 0.4295 & 505.6 & 42.55 & 51.73 & 1.66 \\ 
\colrule
\end{tabular}
\end{table}

Of the two solutions obtained with 24 channels, 
the one with the larger radius gives a much better fit to the data
with $\chi^2/N$ = 1.66 and the associated potential will be referred 
to as the M3Y+rep-2 potential. The potential for the solution with 
the smaller radius is called the M3Y+rep-1 potential.
The two entrance channel potentials are illustrated in Fig. 1. 
An important question is whether the parameters of the stable 
solutions are realistic, or whether some of them can be ruled out
as being unrealistic. One parameter of particular interest is the
radius which is examined below.


The rms (root-mean-square)
radii obtained for the stable solutions are shown in the 
fourth column of Table \ref{m3yradi}. They can be compared to the
estimated experimental rms matter radius of $^{48}$Ca
shown in the last line of the Table.
The estimate was based on the rms radius of the proton distribution,
which was obtained from the measured rms charge radius \cite{angeli},
and the experimental rms radius of the neutron distribution \cite{ray}.
The neutron radius is uncertain and several values exist in the 
literature. The experimental value chosen here was obtained from an
analysis of elastic proton scattering data at 800 MeV \cite{ray} and 
is in fairly good agreement with most of the theoretical predictions 
shown in Table 1 of Ref. \cite{dobi}.

\begin{table}
\caption{The radius $R$ of $^{48}$Ca extracted from
the analysis of the fusion data; c.~f. Table \ref{xki4848}. 
The rms radii are compared to the measured rms charge \cite{angeli}, 
proton \cite{angeli}, and neutron \cite{ray} radii. 
The latter two have been combined into
the rms matter radius shown in the last line. 
The quoted matter-radius ([3.75] fm) was derived by inserting 
the rms matter-radius and the diffuseness $a$ = 0.54 fm 
into Eq. (\ref{rms}).}
\label{m3yradi}
\begin{tabular} {|c|c|c|c|}
\colrule 
Reference & No of Ch. & $R$ (fm) & $\langle r^2\rangle^{1/2}$ (fm) \\
\colrule 
          & 9   & 3.755 & 3.547 \\ 
          & 9   & 3.810 & 3.569 \\ 
M3Y+rep-1 & 24 & 3.745 & 3.528 \\ 
M3Y+rep-2 & 24 & 3.798 & 3.562 \\ 
\colrule 
Charge  \cite{angeli} & &         & 3.474(1) \\ 
protons                    & &         & 3.387(1) \\ 
neutrons   \cite{ray}      & &         & 3.63(5)  \\
matter                     & & [3.75]  & 3.53(3) \\
\colrule 
\end{tabular}
\end{table}

The estimated rms matter radius quoted in the last line of Table 
\ref{m3yradi} is in perfect agreement with the rms radius associated 
with the M3Y+rep-1 solution. The rms radius for the M3Y+rep-2 solution 
is larger but it is still consistent with the experimental estimate
within the 1$\sigma$ uncertainty. A possible explanation for the
larger radius could be the influence of the polarization of
high-lying states not included in the calculations (see below.)  

It is also encouraging that the extracted values of $a_r$ 
shown in Table \ref{xki4848} are similar to those 
determined in the analysis of the fusion data for $^{64}$Ni+$^{64}$Ni 
($a_r$ = 0.403 fm \cite{misi75}), $^{16}$O+$^{16}$O ($a_r$ = 0.41 fm 
\cite{o16}) and $^{48}$Ca+$^{96}$Zr ($a_r$ = 0.40 fm \cite{ca48zr}.) 
It is noted that in the previous works the densities (including the 
radius) were kept fixed and only the value of $a_r$ was adjusted in 
each case to improve the fit to the data.


\subsection{Polarization effects}

The polarization effect discussed in the introduction section is 
illustrated in Table \ref{xki2de}. The Table shows that for
the M3Y+rep-2 potential, one needs to shift the 1 channel 
calculation by $\Delta E$ = -0.80 MeV and the 9 channel calculation 
by -0.12 MeV in order to optimize the fit to the data. 
The negative energy shifts are equivalent to using a larger radius 
of the reacting nuclei. For example, the required energy shift of 
-0.12 MeV for the calculations with 9 channels can be simulated by 
increasing the radius of $^{48}$Ca by only 0.02 fm. 

The required energy shift $\Delta E$ shown in Table \ref{xki2de}
for the calculation with 24 channels is zero simply because the radius $R$ 
was already adjusted in this case  to optimize the fit to the data. 
The issue whether the calculations have converged with respect to the 
excitation and polarization of high-lying states is a difficult 
question to answer. It is possible that the polarization of other 
high-lying states, which have not been considered here, could 
play a role and explain part of the 0.05 fm difference between the 
radius of the M3Y+rep-2 solution and the estimated experimental 
matter radius (see Table \ref{m3yradi}.)

\begin{table}
\caption{The $\chi^2/N$ (column 3) obtained in calculations that use 
the M3Y+rep-2 potential and include 1, 9 or 24 channels.
The energy shift $\Delta E$ of the calculations that optimizes
the fit to the data, and the associated $\chi^2/N$, are shown in 
the last two columns.}
\label{xki2de}
\begin{tabular} {|c|c|c|c|c|}
\colrule
 $R$ (fm) & channels & $\chi^2/N$ & $\Delta E$ (MeV) & $\chi^2/N$ \\
\colrule
 3.798 &  1 & 33.3  & -0.80  & 4.65 \\
 3.798 &  9 &  4.28 & -0.12  & 2.71 \\
 3.798 & 24 &  1.66 &  0.0   & 1.66 \\
\colrule
\end{tabular}
\end{table}


\subsection{$S$ factor representation}

\begin{figure}
\includegraphics[width = 8cm]{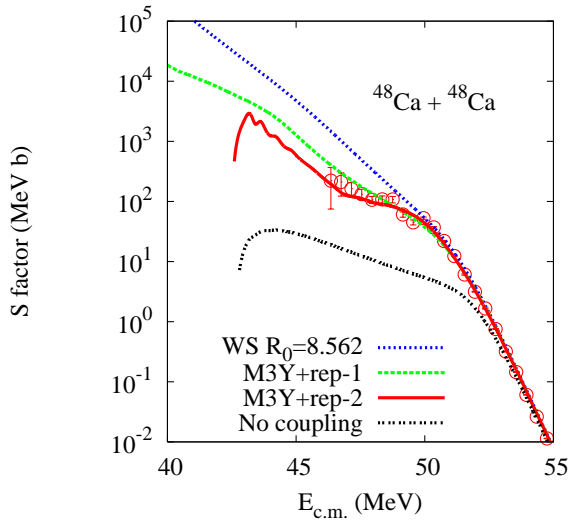}
\caption{\label{m3y4848sf} (Color online) 
$S$ factors for the fusion cross sections shown in Fig. \ref{m3y4848ff}.
Also shown is the $S$ factor obtained in calculations with the
M3Y+rep-1 potential. All calculations include 24 channels,
except the no-coupling limit which has only 1 channel.}
\end{figure}

A good way to illustrate the behavior of the fusion cross section 
$\sigma_f$ at low energies is to plot the $S$ factor for fusion,
\begin{equation}
S = E_{c.m.} \ \sigma_f \ \exp(\eta-\eta_0),
\end{equation}
where $\eta=Z_1Z_2e^2/(\hbar v)$ is the Sommerfeld parameter and 
$\eta_0$ is that value of $\eta$ at a fixed reference energy $E_0$. 
The $S$ factors for the fusion cross sections shown in 
Fig. \ref{m3y4848ff} are illustrated in Fig. \ref{m3y4848sf} 
using the (arbitrary) reference energy $E_0$ = 52 MeV.
Also shown is the result obtained with the M3Y+rep-1 potential 
and 24 coupled channels.

The coupled-channels calculations for the Woods-Saxon potential 
produce an $S$ factor in Fig. \ref{m3y4848sf} that keeps increasing 
with decreasing energy.
The $S$ factors obtained with the two M3Y+repulsion potentials 
and 24 coupled channels are lower.
The $S$ factor for best fit to the data (the solid curve, obtained 
with the M3Y+rep-2 potential) has a maximum at $E_s$ = 43.2 MeV. 
The cross section associated with the latter maximum is very small, 
about 0.3 nb.  The $S$ factor for the calculation based on the 
M3Y+rep-1 potential has a maximum at $E_s$ = 35.4 MeV which is 
outside the depicted energy range.

It would be very interesting to know whether the predicted $S$ factor 
maximum near $E_s$ = 43.2 MeV can be confirmed by experiments but to 
measure a cross section of only 0.3 nb would be a serious challenge.

\subsection{Logarithmic derivative}

\begin{figure}
\includegraphics[width = 8cm]{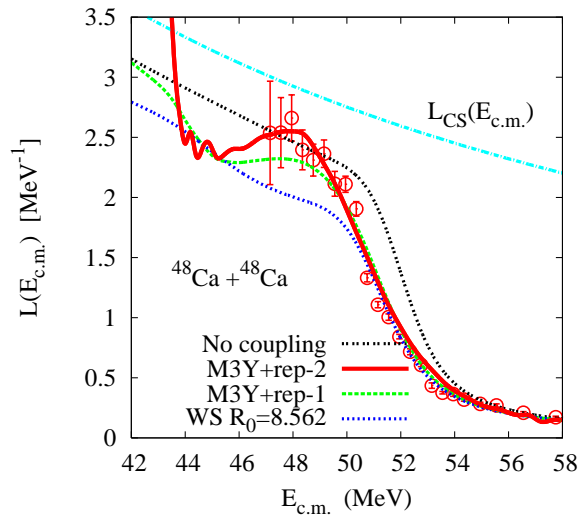}
\caption{\label{4848f1d} (Color online) 
The logarithmic derivative of the energy weighted cross sections
shown in Fig. \ref{m3y4848ff}.
Also shown is the result of coupled-channels calculations based 
on the M3Y+rep-1 potential.
The top curve is the constant $S$ factor limit, $L_{CS}(E_{c.m.})$.} 
\end{figure}

The logarithmic derivative of the energy-weighted cross sections, 
\begin{equation}
L(E_{c.m.}) = \frac{d}{dE_{c.m.}}
\ln\Bigl(E_{c.m} \sigma_f\Bigr),
\end{equation}
is illustrated in Fig. \ref{4848f1d}. The logarithmic derivatives 
derived from the data and from the coupled-channels calculation 
based on the M3Y+rep-2 potential (the solid curve) 
are seen to be in very good agreement.
Similar results were obtained in Ref. \cite{stef4848} in coupled-channels 
calculations that used a Woods-Saxon potential with the large diffuseness
$a$ = 0.9 fm. Thus it appears that a large diffuseness of the 
Woods-Saxon has an effect that is similar to that of the shallow 
M3Y+repulsion potential, at least in the low-energy region
discussed here. 

The similarity of the M3Y+repulsion and a Woods-Saxon potential 
was recently pointed out in Ref. \cite{godsi}. It was shown that 
the M3Y+repulsion potential can be reproduced accurately in the 
barrier region by a Woods-Saxon potential with large diffuseness.
However, a nuclear potential with a large diffuseness is inconsistent
with many measurements of elastic and quasielastic scattering.
For example, a recent systematic study of the quasielastic 
scattering of nuclei showed that a realistic diffuseness in the 
range of 0.64 to 0.69 fm is indeed required \cite{cjlin}.

It is very interesting to point out that the low-energy behavior of 
the experimental logarithmic derivative shown in Fig. \ref{4848f1d}
is different from the behavior observed in other systems, in particular 
in medium-heavy systems \cite{jiangsys}, where the logarithmic derivative 
usually increases linearly with decreasing energy and intersects with 
the logarithmic derivative for constant $S$ factor \cite{jiang2004},
\begin{equation}
L_{CS}(E_{c.m.})=\pi\eta/E_{c.m.}. 
\end{equation}
However, there are other systems that exhibit a deviant behavior at low 
energies. For example, the logarithmic derivative for the fusion of 
$^{36}$S+$^{48}$Ca \cite{sca48} also becomes rather flat at low energies 
and it seems unlikely it will intersect with the constant $S$ factor limit. 
In fact, the $S$ factor for the fusion of $^{36}$S+$^{48}$Ca increases
slowly and linearly in a logarithmic plot with decreasing energy
(see Fig. 3 of Ref. \cite{sca48}.)
  
Another interesting point is that the solid curve in Fig. \ref{4848f1d}
exhibits a maximum near 48 MeV before it rises steeply below 44 MeV.
It intersects with the constant $S$ factor limit at $E_s$ = 43.2 MeV,
where the associated $S$ factor in Fig. \ref{m3y4848sf} develops a maximum. 


\section{Conclusions}

The fusion data for $^{48}$Ca+$^{48}$Ca have been analyzed using
the coupled-channels technique and different ion-ion potentials. 
The analysis based on a standard Woods-Saxon potential clearly 
showed that the data are strongly hindered at low energies. 
By employing and adjusting the M3Y+repulsion double-folding potential
it was possible to achieve an excellent description of the data. 

The best fit to the data was achieved with a nuclear radius of $^{48}$Ca
that is slightly larger than but still consistent with the matter radius 
of $^{48}$Ca. The latter radius was determined from the measured rms 
charge radius and the rms neutron radius extracted from an analysis 
of elastic proton scattering data.
The fact that the extracted radius is slightly larger than the 
matter radius may be caused by the polarization of high-lying states 
that are not included in the coupled-channels calculations.

The entrance channel potential for the best fit to the data has a rather 
shallow pocket, consistent with the findings of previous analyses of 
fusion data for medium-heavy systems. 
The M3Y+repulsion potential model is therefore also referred to as 
the shallow potential model, in contrast to models based on the 
standard Woods-Saxon potentials, which have relatively deep pockets 
in the entrance channel potential.

The $S$ factor for the fusion of $^{48}$Ca+$^{48}$Ca does not show
a maximum within the energy range of the experiment. However, it is 
predicted to develop a maximum at a 3 MeV lower energy which is nearly 
the same as the energy value obtained from the extrapolation method 
in Ref. \cite{jiang4048}.
The cross section associated with the maximum $S$ factor is very small 
($\approx$ 0.3 nb) and is a serious challenge to the experimental
technology.

{\bf Acknowledgments}. 
One of the authors (H.E.) acknowledges discussions with 
\c S. Mi\c sicu about double-folding potentials. 
This work was supported by the U.S. Department of Energy,
Office of Nuclear Physics, contract no. DE-AC02-06CH11357.

\end{document}